\documentclass[aps,prl,reprint,superscriptaddress]{revtex4-1}
\usepackage{amsmath,bm,amssymb,amsfonts}
\usepackage{dcolumn}
\usepackage{graphicx}
\usepackage{latexsym}
\usepackage{epsfig}
\usepackage{multirow}

\usepackage{color}

\begin{document}

\title{Orbital-selective Mott phase and non-Fermi liquid in FePS$_3$}

\author{Minsung \surname{Kim}}
\affiliation{Department of Physics and Astronomy, Rutgers University, Piscataway, New Jersey 08854, USA}
\author{Heung-Sik \surname{Kim}}
\affiliation{Department of Physics and Institute for Accelerator Science, Kangwon National University, Chuncheon 24341, Korea}
\author{Kristjan \surname{Haule}}
\affiliation{Department of Physics and Astronomy, Rutgers University, Piscataway, New Jersey 08854, USA}
\author{David \surname{Vanderbilt}}
\affiliation{Department of Physics and Astronomy, Rutgers University, Piscataway, New Jersey 08854, USA}

\date{\today}

\begin{abstract}
The layered metal phosphorous trisulfide FePS$_3$ is reported to be a Mott insulator at ambient conditions and to undergo structural and insulator-metal phase transitions under pressure.
However, the character of the resulting metallic states has not been understood clearly so far.
Here, we theoretically study the phase transitions of FePS$_3$ using first-principles methods based on density functional theory and embedded dynamical mean field theory.
We find that the Mott transition in FePS$_3$ can be orbital-selective, with $t_{2g}$ states undergoing a correlation-induced insulator-to-metal transition while $e_g$ states remain gapped.
We show that this orbital-selective Mott phase, which occurs only when non-hydrostatic pressure is used, is a bad metal (or non-Fermi liquid) with large fluctuating moments due to Hund's coupling.
Further application of pressure increases the crystal-field splitting and converts the system to a conventional Fermi liquid with low-spin configurations dominant.
Our results show that FePS$_3$ is a novel example of a system that realizes an orbital-selective Mott phase, allowing tuning between correlated and uncorrelated metallic properties in an accessible pressure range ($\leq$ 18 GPa). 
\end{abstract}


\date{\today}

\maketitle

\emph{Introduction}---
The Mott transition is a prototypical manifestation of correlation effects in the electronic structure of materials in which theoretical considerations beyond conventional band theory become essential for the proper description of the electron localization~\cite{mott1937,mott1949,hubbard1963,imada1998}. In principle, electron correlation effects can be orbital-dependent in the sense that the critical strength of the correlation required for the localization can vary for different orbitals belonging to ``decoupled'' manifolds. In such a case, some of the electrons become localized due to the Mott transition, while others remain metallic near the Fermi energy. This orbital-selective Mott transition (OSMT) has been discussed and studied in a number of models and materials~\cite{georges2013,anisimov2002,nakatsuji2003,koga2004,biermann2005,liebsch2005,laad2006,werner2007,craco2009,demedici2009,huang2012,herbrych2018,herbrych2020}. However, real material systems with tunable property across the phase boundary under moderate changes of external parameters (e.g., pressure) are rare in the literature.

The layered metal phosphorous chalcogenide FePS$_3$ has been reported to have an insulator-to-metal transition (IMT) and two structural phase transitions separating structural phases HP0, HP1, and HP2 under pressure~\cite{haines2018,coak2019}. The ambient-pressure phase HP0 is known to be a Mott insulator while the highest-pressure phase HP2 is metallic. The intermediate pressure phase HP1 was reported to be metallic based on the temperature dependence of the resistivity in single-crystal transport measurements of Ref.~\cite{haines2018}, whereas it was assigned to be gapped later in Ref.~\cite{coak2019}. This apparent contradiction can be attributed to different pressure conditions. The former experiment was performed under non-hydrostatic (or quasi-hydrostatic) pressure, arising from the use of a powder pressure-transmitting medium, while the pressure was effectively hydrostatic in the latter~\cite{coak2020commun,haines2018,coak2019}. The non-hydrostatic condition, with larger pressure component in the direction normal to the plane than in-plane, was essential to realize the metallic HP1 state~\cite{coak2020commun}. A notable feature under non-hydrostatic pressure is that the resistivity of the metallic HP1 phase is a few orders of magnitude larger than that of HP2. This suggests that the HP1 phase could be a ``bad metal'' phase since the resistivity tends to increase as temperature $T$ increases in the high-$T$ regime in HP1 as well as HP2. Indeed, we find that HP1 shows a bad metal behavior due to the correlation-induced OSMT, as will be discussed in detail below. This is a distinct feature of FePS$_3$ compared to other compounds such as MnPS$_3$ and NiPS$_3$ in the same material class~\cite{wang2016,sykim2018,harms2020,hskim2019,neal2020}, which do not show the OSMT.

Here, we theoretically investigate the phase transitions in FePS$_3$ under pressure using first-principles methods based on a combination of density functional theory (DFT) and embedded dynamical mean field theory (eDMFT)~\cite{blaha2020,haule2010,haule2015,haule2016,haule2007,kotliar2006,georges1996}. Most importantly, we find that the metallic HP1 phase in Ref.~\cite{haines2018} is an orbital-selective Mott phase (OSMP) with $t_{2g}$ ($=a_{1g}+e_g^\prime$) states becoming metallic while $e_g$ states remain gapped. This novel feature of the experimentally-realized system has remained unnoticed so far. Also, there has been no theoretical understanding of the origin of the large resistivity difference in the two metallic phases HP1 and HP2. Our calculations show that the OSMP of HP1 is a non-Fermi-liquid phase with bad metallic behavior while HP2 is a conventional Fermi-liquid. We find that the key element for the qualitatively different metallic phases is the competition between the Hund's coupling and the crystal-field splitting. Our theory further clarifies the relation between the structural and electronic phase transitions and the effect of non-hydrostatic pressure conditions, in good agreement with experiments~\cite{haines2018,coak2019}. Thus, FePS$_3$ presents an intriguing example of a correlated system where three contrasting phases (i.e., Mott insulator, non-Fermi liquid, and Fermi liquid) appear in an accessible pressure range ($\leq$ 18 GPa).

\emph{Theoretical methods}---
To study the structural phase transitions under pressure, we performed DFT calculations as implemented in Vienna {\it Ab initio} Simulation Package (VASP)~\cite{kresse1993,kresse1996}. The projector augmented-wave (PAW) method was used to describe the interaction between ions and valence electrons~\cite{blochl1994,kresse1999}. We employed a plane-wave basis set with a 516 eV energy cutoff and used $8\times6\times4$ (for HP0 and HP1 structures) and $8\times8\times10$ (for HP2) $k$-point grids. The Perdew-Burke-Ernzerhof exchange-correlation functional~\cite{perdew1996} was used. We adopted $U_{eff}$=2.5 eV for the DFT+U effective on-site Coulomb repulsion, which was reported to reproduce the experimental structural phase transitions under pressure in FePS$_3$~\cite{zheng2019}. The van der Waals (vdW) energy was accounted for using the DFT-D2 approach~\cite{grimme2010}. The atomic positions were relaxed until the residual forces became $\leq$ 0.01 eV/\AA.
To study the electronic phase transitions, first-principles calculations based on the combination of DFT and eDMFT as implemented in \texttt{WIEN2k} and the Rutgers eDMFT code were performed~\cite{blaha2020,haule2010}. We set $RK_{max}$ (which determines the size of basis) to be 7.0 and used 500 $k$-points for the $k$-point sampling in the Brillouin zone. We adopted the local-density approximation (LDA)~\cite{perdew1992}, which gives the best results for lattice properties when combined with eDMFT~\cite{haule2015}. The atomic positions were relaxed with the force tolerance 2\,mRy/Bohr in paramagnetic configurations at $T$ = 300 K using eDMFT~\cite{haule2016}, while the lattice parameters were fixed to the values obtained from the VASP calculation at each pressure. We adopted $U$=8.0 eV and $J_H$=0.8 eV for the Coulomb repulsion and Hund's coupling respectively. The auxiliary impurity problem was solved using a continuous-time quantum Monte Carlo impurity solver~\cite{haule2007}.
Since we are interested in the metallic HP1 phase in this study, the focus will be the non-hydrostatic pressure hereafter. We notice in passing that the hydrostatic pressure results in Mott insulating HP1 phase, which is transformed through the first order structural transition to the conventional HP2 metallic phase under higher pressures, such that the OSMP in that case is hidden in experiment~\cite{coak2019}. We consider a larger out-of-plane pressure component than in-plane as in the experiment~\cite{haines2018}. Specifically, we chose $P_{zz}$ = 1.4 $P$ and  $P_{xx}$ = $P_{yy}$ = 0.8 $P$ with $P = (P_{xx}+P_{yy}+P_{zz})/3$ and also tested other values of the anisotropy factor $P_r$ = $P_{zz}/P$ (see the Supplemental Material~\cite{suppmater}).

\begin{figure}[t]
\includegraphics[width=0.48\textwidth]{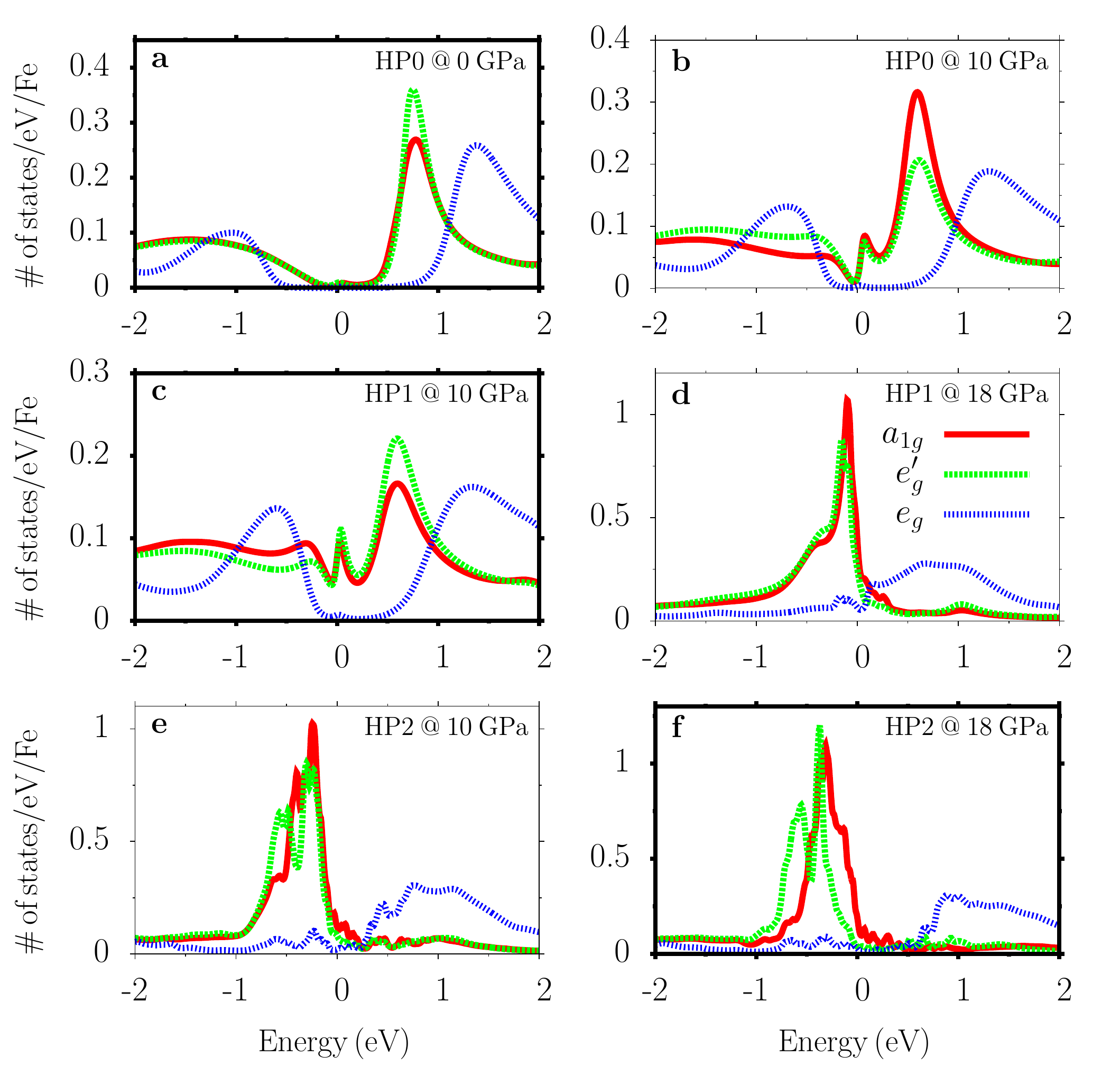}
\caption{\label{fig:pdosstr} Electronic structures with or without expected structural changes under pressure. The partial densities of states are depicted for (a) HP0-SPD at 0 GPa, (b) HP0-SPD at 10 GPa, (c) HP1-APD at 10 GPa, (d) HP1-APD at 18 GPa, (e) HP2-APC at 10 GPa, and (f) HP2-APC at 18 GPa. Figures (a), (c), and (f) are for stable configurations at the given pressure and are denoted by thicker borders, whereas (b), (d), and (e) are fictitious metastable structures for comparison. The red solid, green dotted, and blue dotted lines correspond to $a_{1g}$, $e_g^\prime$, and $e_g$ states, respectively, as denoted in (d). The Fermi level is set to zero.
}
\end{figure}

\begin{figure}[]
\includegraphics[width=0.48\textwidth]{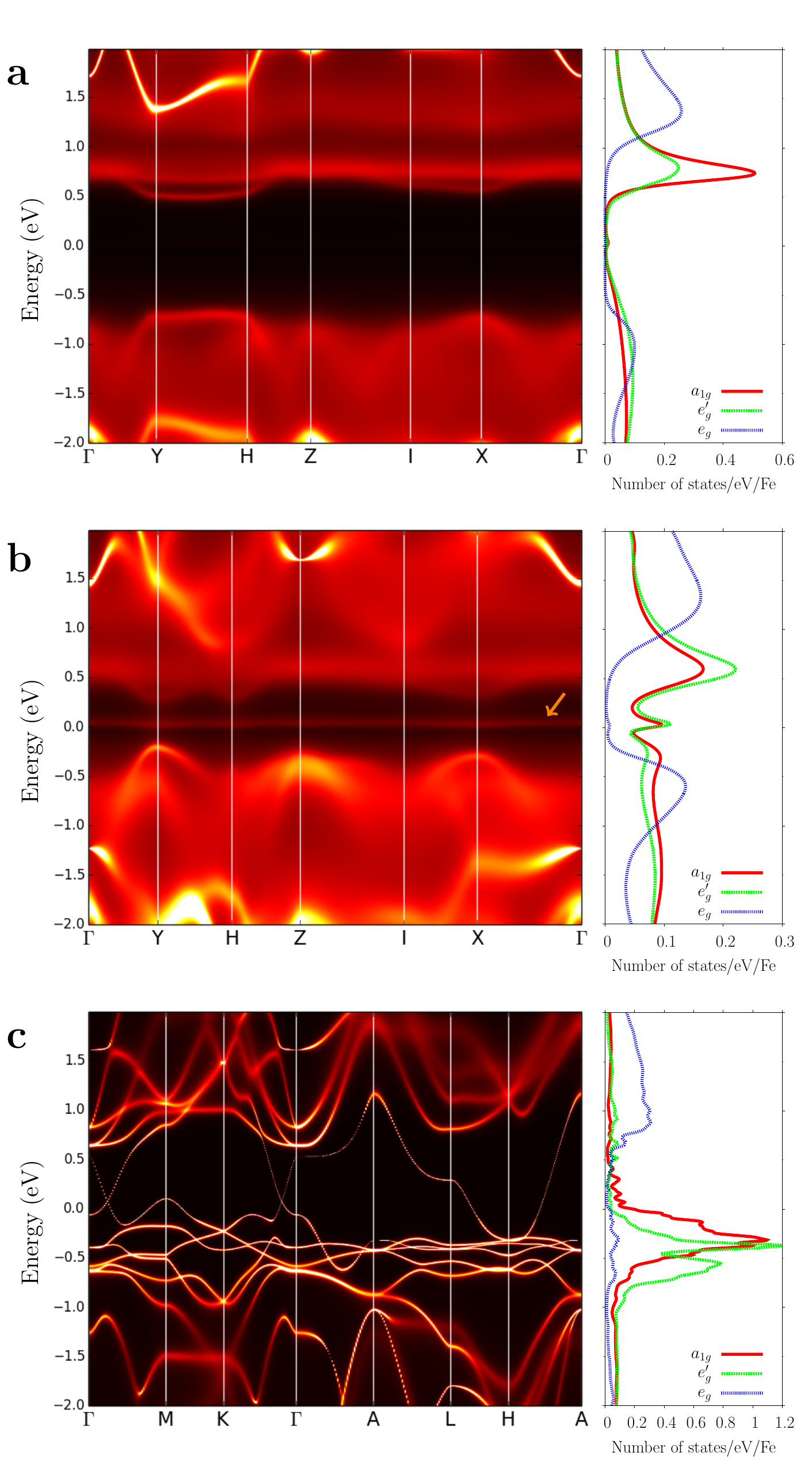}
\caption{\label{fig:sfpdos} $k$-resolved spectral functions and PDOS for the 3 different phases at representative pressure values under the non-hydrostatic pressure condition. $A(k,\omega)$ and the corresponding PDOS are plotted for (a) HP0-SPD (0 GPa), (b) HP1-APD (10 GPa), and (c) HP2-APC (18 GPa).
}
\end{figure}

\emph{Intertwined electronic and structural phase transitions}---
The structural phase transitions play an important role in the realization of the different electronic phases of FePS$_3$ under pressure. The three structural phases of FePS$_3$ can be characterized by the arrangement of P atoms, and the two structural phase transitions among HP0-SPD (staggered P dimers), HP1-APD (aligned P dimers), and HP2-APC (aligned P chains) are reported to be induced by layer sliding at P$_{c1} \approx$ 4 GPa and inter-layer collapse in vdW gaps at P$_{c2} \approx$ 14 GPa~\cite{haines2018,coak2019,zheng2019}. This is confirmed by our DFT calculations (for details of the DFT calculations regarding the structural properties, see the Supplemental Material~\cite{suppmater}). 

The ambient pressure phase HP0-SPD is a Mott insulator as can be seen by the vanishing projected density of states (PDOS) and spectral weight at the Fermi energy (Fig.~\ref{fig:pdosstr}a and Fig.~\ref{fig:sfpdos}a). On the other hand, HP1-APD at 10 GPa shows metallic behavior with finite spectral weight and DOS at the Fermi level. Notably, the IMT occurs only in the $t_{2g}$ (i.e., $a_{1g}+e_g^\prime$) sector, whereas the $e_g$ states remain gapped (Fig.~\ref{fig:pdosstr}c and Fig.~\ref{fig:sfpdos}b). Here, we note that the layer sliding increases the hybridization of $t_{2g}$ orbitals and makes them metallic, since HP0-APD at 10 GPa still has a small gap (Fig.~\ref{fig:pdosstr}b). This OSMP has incoherent metallic states derived from $t_{2g}$ states near the Fermi energy (indicated by an orange arrow in Fig.~\ref{fig:sfpdos}b). These metallic states show non-Fermi-liquid character, suggesting that the magnetic moments of the localized electrons act as scattering centers for the itinerant electrons~\cite{demedici2009}. The highest pressure phase HP2-APC at 18 GPa also shows metallic features (Fig.~\ref{fig:pdosstr}f). However, in contrast to the HP1-APD, the metallic states in HP2-APC show clear quasi-particle peaks near the Fermi energy (Fig.~\ref{fig:sfpdos}c), confirming conventional Fermi liquid behavior. Thus, beyond the pressure-induced IMT, we find two qualitatively different metallic states. Here, the inter-layer collapse disfavors the OSMP as can be seen in HP2-APC at 10 GPa (Fig.~\ref{fig:pdosstr}e). A sufficiently large pressure in the HP1-APD (Fig.~\ref{fig:pdosstr}d) increases the crystal field strength and also disfavors the OSMP. 

Finally, the non-hydrostatic condition is important for the realization of the OSMT. We find that larger anisotropy in the pressure condition (i.e., larger out-of-plane pressure component compared with in-plane ones) is advantageous to the metallization of $t_{2g}$ states, while we find a small conventional Mott gap under the hydrostatic pressure (see the Supplemental Material~\cite{suppmater}). The non-Fermi liquid character and the effect of the crystal fields in comparison with the Hund's coupling will be discussed in more detail below.

\begin{figure}[]
\includegraphics[width=0.48\textwidth]{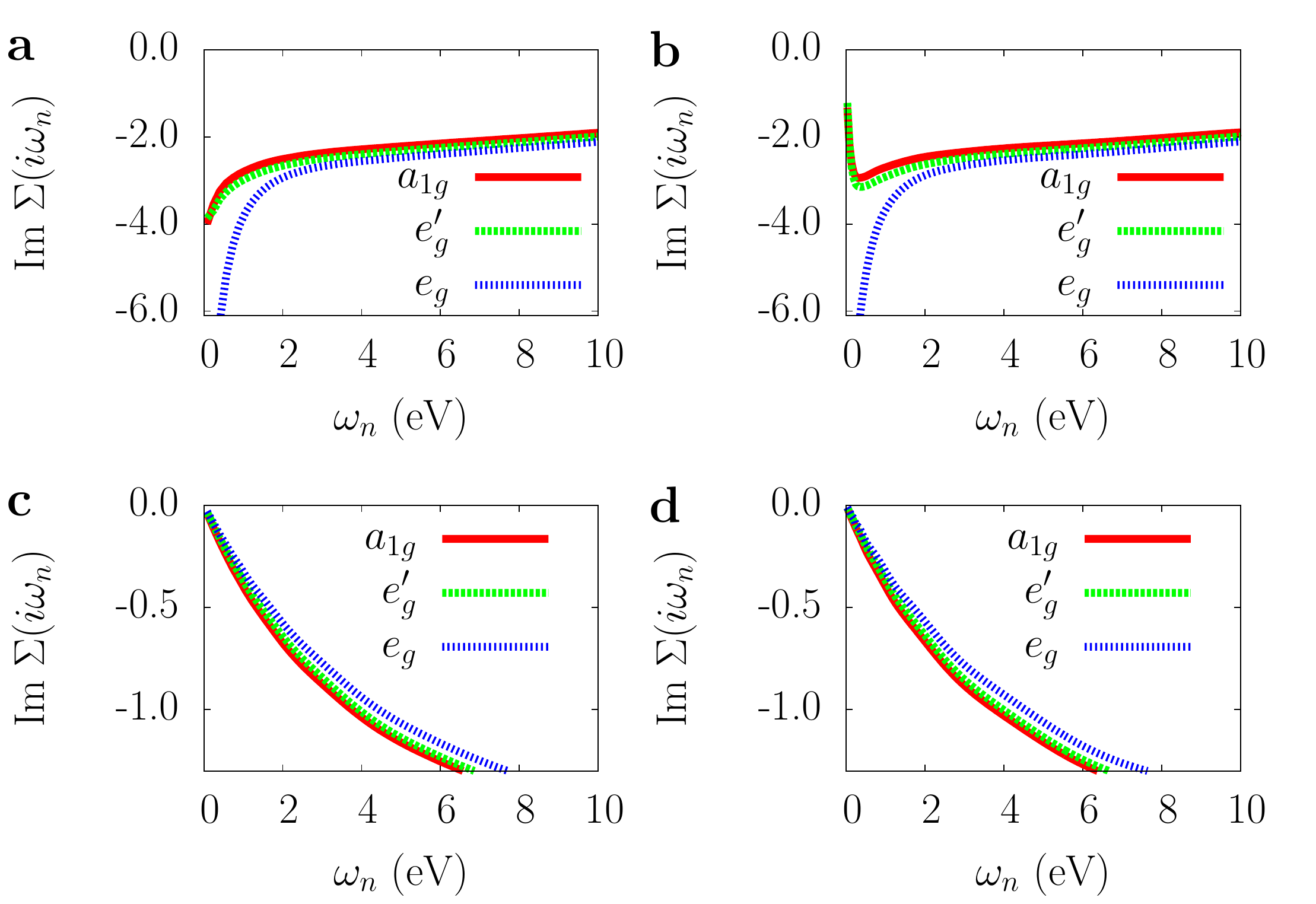}
\caption{\label{fig:imsig} Imaginary part of self-energy on the imaginary axis. Im $\Sigma(i\omega_n)$ for HP1-APD (10 GPa) at (a) 300 K and (b) 100 K, and for HP2-APC (18 GPa) at (c) 300 K and (d) 100 K.
}
\end{figure}

\emph{Non-Fermi liquid vs. Fermi liquid}---
To further understand the two contrasting metallic states, we examine the scattering rate $\Gamma \sim {\rm Im}~\Sigma(i0^+)$ where $\Sigma$ is the self-energy. For Fermi liquids, the scattering rate is supposed to behave as $\sim T^2$ due to the small phase space for scattering in the conventional Fermi liquid, and $\Sigma(i0^+)$ would become negligible at sufficiently low $T$. Thus, by inspecting the behavior of  $\Sigma(i0^+)$ at different $T$, one can examine whether the system is close to a Fermi liquid. In Fig.~\ref{fig:imsig} we compare $\Sigma(i\omega_n)$ for two different temperatures ($T$ = 300 K and 100 K). For the HP1-APD at 10 GPa, we find that $\Sigma(i0^+)$ is reduced for $t_{2g}$ states at the low $T$ as expected, but still remains large at $T$ = 100 K. By contrast, $\Sigma(i0^+)$ is very small at both temperatures for HP2-APC at 18 GPa. The large scattering rate of the metallic states in HP1-APD is attributed to fluctuating magnetic moments of the localized ($e_g$ in our case) electrons. Due to the large scattering rate, the OSMP in HP1-APD shows a bad-metal non-Fermi-liquid behavior with higher resistivity compared with conventional metals~\cite{biermann2005,liebsch2005}. 
Also, at low $T$ the magnitude of $\Sigma(i\omega_n)$ of $t_{2g}$ decreases rapidly as $\omega_n \rightarrow 0$ in HP1-APD (Fig.~\ref{fig:imsig}b), whereas that of Fermi liquids decreases linearly as in HP2-APC (Fig.~\ref{fig:imsig}d). This rapid decrease is also discussed in model Hamiltonian studies of the OSMP~\cite{demedici2009,biermann2005,werner2006}, where logarithmic~\cite{biermann2005} and power-law~\cite{werner2006} behaviors have been reported.

\begin{figure}[b]
\includegraphics[width=0.48\textwidth]{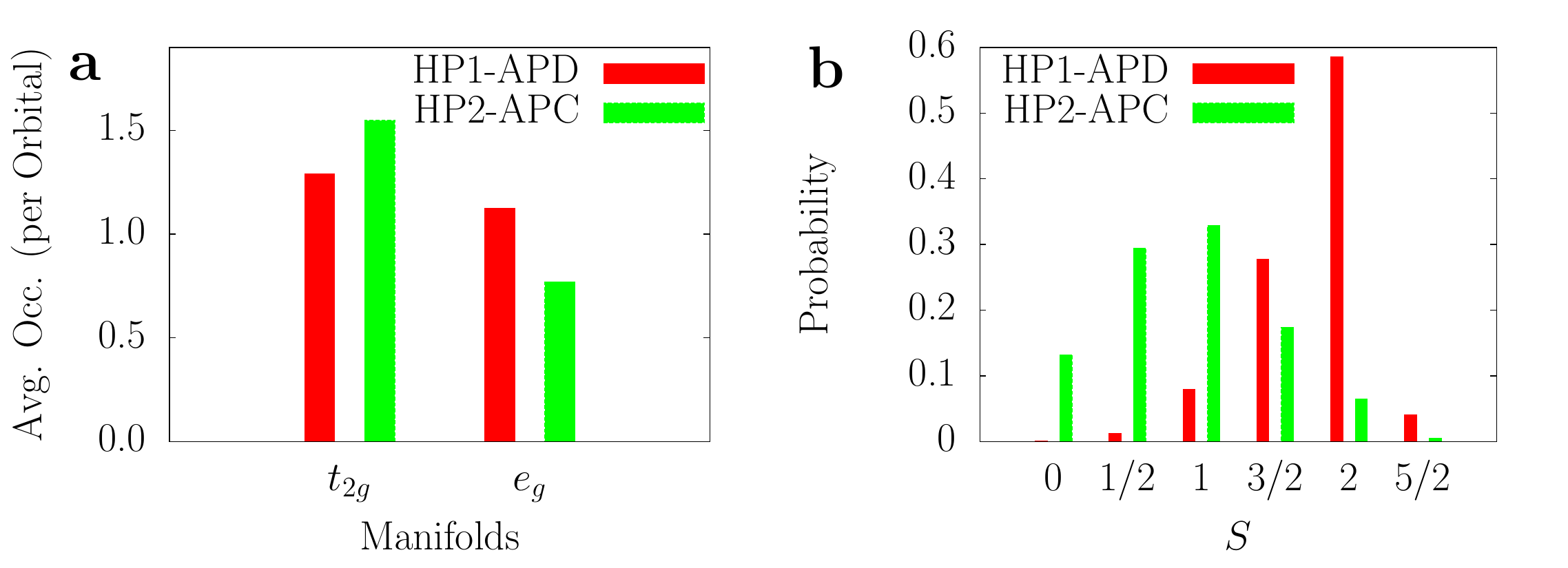}
\caption{\label{fig:occspin} Average occupation and spin configurations. (a) Occupation per orbital and (b) probability of spin configurations in HP1-APD (10 GPa) and HP2-APC (18 GPa).
}
\end{figure}

\emph{Hund's coupling vs. Crystal field splitting}---
The different relative strengths of the Hund's coupling and the crystal field splitting give rise to contrasting spin configurations in HP1-APD and HP2-APC (Fig.~\ref{fig:occspin}). When the Hund's coupling is dominant as in HP1-APD, high-spin configurations are favored, where the occupation in each orbital in the $t_{2g}$ and the $e_g$ manifolds is similar (see the PDOS in Fig.~\ref{fig:sfpdos}b and the occupation per orbital and spin configurations in Fig.~\ref{fig:occspin}). On the other hand, if the crystal field is large as in HP2-APC, low-spin configurations are favored. Since the Fe ions in FePS$_3$ are nominally Fe$^{2+}$ with 6 electrons in the 3$d$ shell, the electrons will occupy the $t_{2g}$ manifold considerably more than the $e_g$ (Fig.~\ref{fig:pdosstr}f and Fig.~\ref{fig:occspin}). As expected from the comparison between the two phases, the sizable strength of Hund's coupling is essential to realize the OSMP. If we artificially set $J_H$ = 0 eV in HP1-APD, the OSMP disappears and the PDOS becomes similar to that of HP2-APC, as confirmed by the relative occupation of the $t_{2g}$ and $e_g$ orbitals (see the Supplemental Material~\cite{suppmater}). 

Also, the Hund's coupling effectively decouples the $t_{2g}$ and $e_g$ band manifolds in HP1-APD. We calculated the orbital fluctuation $\langle (n_{A} - \langle n_{A} \rangle) (n_{B} - \langle n_{B} \rangle) \rangle = \langle n_{A}n_{B} \rangle - \langle n_{A} \rangle \langle n_{B} \rangle$ (i.e., the correlation in the occupation of states A and B) between the $t_{2g}$ and $e_g$ manifolds. We find the inter-manifold fluctuation $\langle n_{t_{2g}}n_{e_g} \rangle - \langle n_{t_{2g}} \rangle \langle n_{e_g} \rangle = -0.075$ in the HP1-APD phase which is markedly smaller in magnitude than the intra-manifold fluctuation in HP1-APD, $\langle n_{a_{1g}}n_{e_g'} \rangle - \langle n_{a_{1g}} \rangle \langle n_{e_g'} \rangle = -0.16$, and the inter-manifold fluctuation in HP2-APC $\langle n_{t_{2g}}n_{e_g} \rangle - \langle n_{t_{2g}} \rangle \langle n_{e_g} \rangle = -0.35$.


The occurrence of the OSMT can be understood qualitatively as follows. If the crystal field is absent, the $t_{2g}$ and the $e_g$ manifolds will have the same energy, making each orbital occupied evenly with $\sim \frac{6}{5}$ electrons. If we turn on the crystal field gradually, the occupation of the $e_g$ states (now higher in energy) will decrease. At some point, the $e_g$ manifold will become half-filled and the OSMT can occur. This scenario is analogous to that of a theoretical study based on a three-band model Hamiltonian with four electrons in Ref.~\cite{demedici2009} in comparison with our five-band system with six electrons.

\begin{figure}[]
\includegraphics[width=0.48\textwidth]{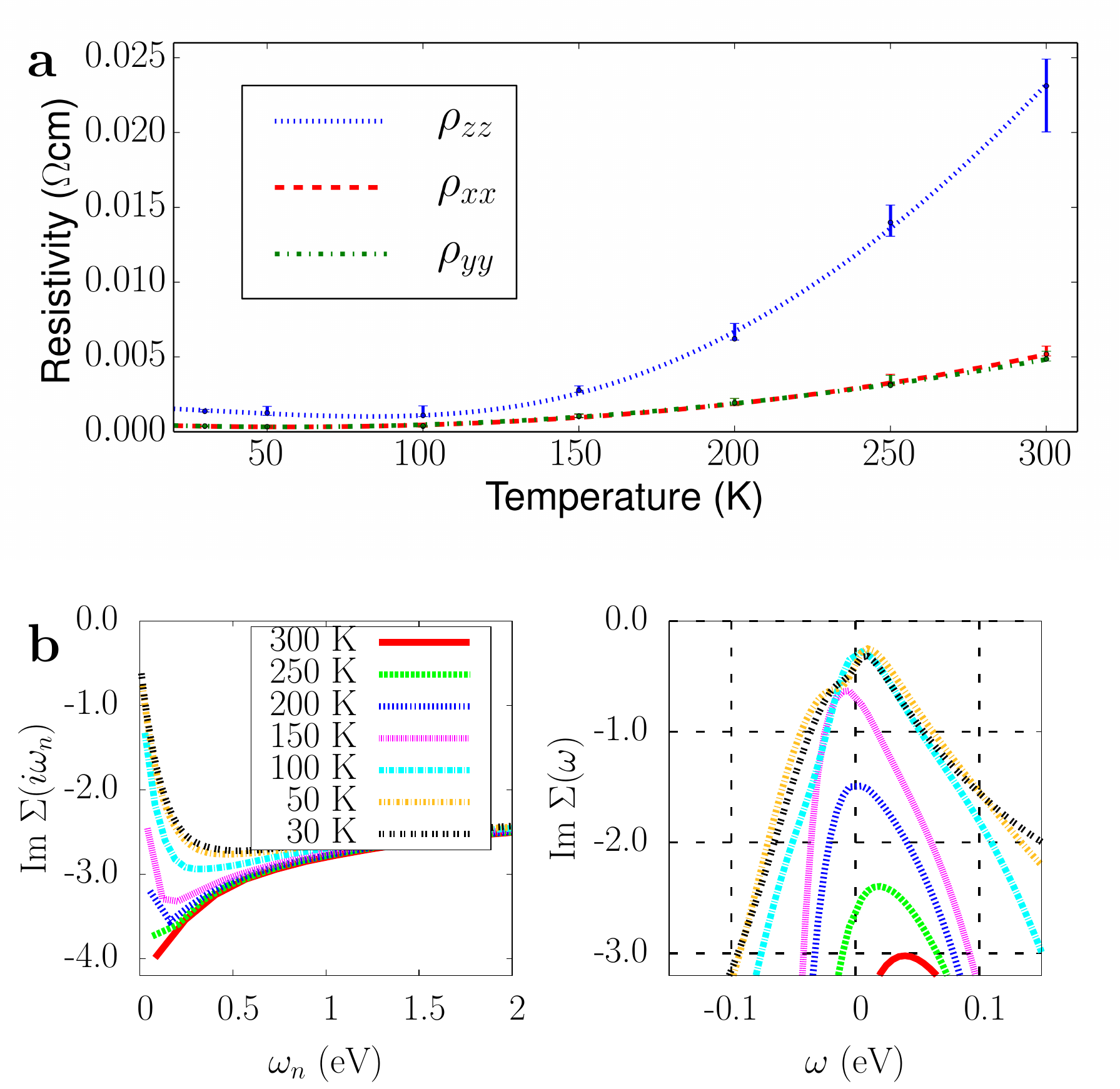}
\caption{\label{fig:resist} Temperature dependence of resistivity. (a) Resistivity for different temperatures. (b) Im $\Sigma$ on (left) the imaginary and (right) the real axes.
}
\end{figure}

\emph{Temperature dependence of resistivity}---
As discussed above, the large scattering rate makes HP1-APD a bad metal. To further study the transport properties of this phase, we calculated the optical conductivity $\sigma(\omega)$ for different $T$, from which we obtained the DC (static) resistivity $\rho(0)$ (Fig.~\ref{fig:resist}a). The resistivity shows metallic behavior in the high-$T$ regime, where it increases as $T$ increases. In Fig.~\ref{fig:resist}b, we show frequency and temperature dependence of the self-energy on the real and imaginary axes, from which the resistivity is computed. As for magnitudes, the resistivity in HP1-APD is about two orders of magnitude larger than that in HP2-APC (in agreement with experiments~\cite{haines2018,coak2019}), as expected from the non-Fermi and the Fermi liquid behavior in the former and the latter phases, respectively. Compared with experiments in the high-T regime~\cite{haines2018,coak2019}, our resistivity values are of the same order of magnitude, but are somewhat smaller, probably due to the lack of other sources of scattering (such as disorder and phonons) in our calculations.


\emph{Conclusion}---
In summary, we investigated the electronic and structural phase transitions in FePS$_3$ under non-hydrostatic pressure. We found that the IMT occurs only in the $t_{2g}$ manifold forming the OSMP, followed by another metal-to-metal transition from a non-Fermi-liquid to a Fermi-liquid state under further application of pressure. The relative strength of Hund's coupling and the crystal-field splitting was important for the realization of the two distinct metallic states. Our study illuminates the salient features of the electronic phase transitions in FePS$_3$ that have been realized experimentally, while their novelty has been overlooked. Our results may be important for the realization of a novel low-dimensional system with tunable correlated electronic properties, and could be useful for the future development of electronic nano-devices.

~\\
\begin{acknowledgments}
We thank Janice L. Musfeldt, Nathan C. Harms, Yexin Feng, and Matthew J. Coak for helpful discussions and comments. This work was supported by NSF DMREF grant (DMR-1629059). H.-S.K. thanks for the support of the National Research Foundation of Korea (Basic Science Research Program, Grant No. 2020R1C1C1005900). This research used resources of the Center for Functional Nanomaterials, which is a U.S. DOE Office of Science Facility, and the Scientific Data and Computing Center, a component of the Computational Science Initiative, at Brookhaven National Laboratory under Contract No. DE-SC0012704.
\end{acknowledgments}

\end{document}